\documentclass[prb,twocolumn,showpacs]{revtex4}

\usepackage{times}
\usepackage{graphicx}
\usepackage{dcolumn}
\usepackage{amsmath}

\newcommand{\LC}{La$_{2/3}$Ca$_{1/3}$MnO$_{3}$ }
\newcommand{\TC}{$T_{\rm C}$}

\begin{document}
\title{Polaronic excitations in CMR manganite films}

\author{Ch. Hartinger}
\affiliation{EP V, Center for Electronic Correlations and
Magnetism, University of Augsburg, 86135 Augsburg, Germany}

\author{F. Mayr}
\affiliation{EP V, Center for Electronic Correlations and
Magnetism, University of Augsburg, 86135 Augsburg, Germany}

\author{A. Loidl}
\affiliation{EP V, Center for Electronic Correlations and
Magnetism, University of Augsburg, 86135 Augsburg, Germany}

\author{T. Kopp}
\affiliation{EP VI, Center for Electronic Correlations and
Magnetism, University of Augsburg, 86135 Augsburg, Germany}

\date{\today}

\begin{abstract}
In the colossal magnetoresistance manganites polarons have been
proposed as the charge carrier state which localizes across the
metal-insulator transition. The character of the polarons is still
under debate. We present an assessment of measurements which
identify polarons in the metallic state of
La$_{2/3}$Sr$_{1/3}$MnO$_{3}$ (LSMO) and
La$_{2/3}$Ca$_{1/3}$MnO$_{3}$ (LCMO) thin films. We focus on
optical spectroscopy in these films which displays a pronounced
resonance in the mid-infrared. The temperature dependent resonance
has been previously assigned to polaron excitations. These
polaronic resonances are qualitatively distinct in LSMO and LCMO
and we discuss large and small polaron scenarios which have been
proposed so far. There is evidence for a large polaron excitation
in LSMO and small polarons in LCMO. These scenarios are examined
with respect to further experimental probes, specifically charge
carrier mobility (Hall-effect measurements) and high-temperature
dc-resistivity.
\end{abstract}


\maketitle

\section{Introduction}

Since the discovery of the colossal magnetoresistance effect in
thin films of \LC by Helmolt \textit{et al.},\cite{Helmolt93} the
properties of doped perovskite manganites have been a focal point
of research in correlated electronic systems. These compounds
reveal a variety of ordinary and exotic ground states, for example
spin, orbital or charge order, cooperative Jahn-Teller (JT)
distortions or electronic phase separation, as a result of a
delicate interplay of orbital, spin, charge, and lattice degrees
of freedom. While in the early days research concentrated on the
appearance of colossal magnetoresistance (CMR), more recently also
issues like orbital order, the importance of electron-phonon
coupling, or phase separation scenarios promoted among the topics
of interest. Optical spectroscopy, in both single crystals and
thin films, has contributed to unravel the complex physics and
provided insight into the importance of electron-phonon coupling
for modelling the optical conductivity in the manganites.

Okimoto {\it et al.}\cite{Okimoto97} and Kaplan {\it et
al.}\cite{Kaplan96}  reported on a large transfer of spectral weight
from high to low energy with decreasing temperature for
La$_{1-x}$Sr$_{x}$MnO$_{3}$ single crystals and for
Nd$_{0.7}$Sr$_{0.3}$MnO$_{3}$ thin films, respectively. The observed
temperature dependence is consistent with the theoretical approach
by Millis \textit{et al.},\cite{Millis96} who went beyond Zener's
double exchange\cite{Zener51} and included dynamic JT polaron
effects. Subsequent studies in
crystalline\cite{Jung98,Kim98,Saitoh99,Lee99} and thin
film\cite{Quijada98,Machida98} samples confirmed the existence of
the mid-infrared (MIR) excitation for a large number of manganites.
In this context the concept of small (SP) and large polarons (LP) as
essential ingredients to understand the electronic transport
properties and the MIR excitation (labeled polaronic peak in the
following) became subject of considerable theoretical
efforts.\cite{Roeder96,Lee97,Millis96,Alexandrov99} In contrast to
an optical transition between electronic bands, a characteristic
feature of a polaronic absorption process is the significant
frequency shift with temperature (see, for example,
Ref.~\onlinecite{Rubinstein99}).

The shape of a polaronic resonance in the optical conductivity
is predicted by theory whereby the details strongly depend on the
nature of the polaronic excitations: large polarons exhibit a
threshold behavior and a coupling dependent
structure\cite{Gurevich62,Goovaerts73} whereas small polarons
display an asymmetric Gaussian peak;\cite{Reik67,Emin75} these
latter polarons may even pair to form bipolarons with their own
characteristics.\cite{Alexandrov99,Zhao00}
The assignment ``large'' or ``small'' polaron usually refers to
the spacial extension of the lattice distortion induced by a single
charge carrier. True to the literature (for example, see the
review by Emin on optical properties of polarons\cite{Emin93})
one would denote a lattice distortion generated by an electronic
charge as a small polaron if it collapses to a single site --- the
smallest state compatible with the atomicity of a
solid. Correspondingly, all polaronic states with extension larger
than a lattice spacing are labelled large polarons. A large polaron
does not necessarily extend over a large number of sites.
However, since we have no direct means of determining the size of
the polaronic cloud from optical spectroscopy, we instead refer to
the related shape of the absorption spectra: threshold behavior
for large polarons versus asymmetric Gaussian peak for small
polarons.\cite{Emin93}

For the metallic manganites, a clear identification of the polaron type and the
respective model is disputable not only because other electronic
correlations may correct or even replace the standard polaronic
physics but also because the statistics of the rather dense
polaron system may strongly influence the shape of the resonance.
Actually, small polarons which are trapped in a locally polarized lattice
are less affected by finite doping. The MIR absorption may still
be visualized in terms of transitions between the adiabatic levels
of neighboring sites. The presence of other localized polarons at
more distant sites will renormalize the polaronic energy scale
(the binding energy) and it will raise the high-frequency wing of
the resonance due to additional multi-polaron processes. However,
a finite density should affect less the low-frequency wing, that
is, absorption at energy less than the binding energy. In this
work we will provide further optical conductivity and
dc-resistivity data which confirm this small polaron scenario for
thin films of La$_{2/3}$Ca$_{1/3}$MnO$_{3}$ (LCMO).

By contrast, a large-polaron resonance should be influenced profoundly by a
finite density of polarons due to their spacial extension and
possible delocalization. A perturbational approach for small
electron-phonon coupling at finite charge carrier density was
presented recently by Tempere and Devreese (TD).\cite{Tempere01}
In fact, if we compare the one-polaron absorption of Gurevich,
Lang and Firsov (GLF)\cite{Gurevich62} to the finite density
result of TD, we notice that the low frequency behavior above the
threshold is dictated by different power laws. Whereas the
one-phonon absorption is characterized by a square root behavior,
the fermionic statistics of many polarons leads to a linear
density of states of particle-hole excitations at the Fermi edge
which is in fact reflected in the threshold behavior of the
absorption. In a recent paper\cite{Hartinger04} we proposed to
identify the MIR resonance in thin films of
La$_{2/3}$Sr$_{1/3}$MnO$_{3}$ (LSMO) by large polarons of this
type.

Here we extend the preliminary discussion of our previous
work in order to elucidate the distinct polaronic behavior in the
metallic phase of manganites with strong (LCMO) and
intermediate-to-weak electron-phonon coupling (LSMO).
The purpose of our discussion is also, to a certain
extent, a critical review of those polaronic models which have
been proposed to identify the resonance in previous measurements
with bulk samples. We observe that hitherto the proposed theoretical models
are not capable of capturing all the aspects of a polaronic metal like
LSMO. Notable details of the  MIR spectra are
in disagreement with the available theoretical models. A more
comprehensive theory which offers an explanation particularly for
the temperature-dependent shift of the polaronic resonance and for the
magnitude of the spectral weight on the high-energy side would probably
clarify the nature of the polaronic metal.

Using thin films in contrast to single crystals, we achieve a lower
conductivity due to grain boundaries and strain.~\cite{Snyder96}
Consequently screening is effectively reduced --- to a level where
phonons and polarons are well observable in the optical
conductivity, even in the metallic phase. This allows the detailed
analysis of phonons\cite{Hartinger04c1,Hartinger04c2} and
polarons\cite{Hartinger04} from the respective optical data.
Besides, the optical properties of single-crystal samples are more
sensitive to surface preparation which, in particular, influences
the MIR-region.\cite{Iida99}

The role of the lattice mismatch on the structural and transport
behavior in dependence of the film thickness has been systematically
investigated in numerous articles. In particular it depends on the
substrate/film combination and on the growth conditions. In general
thicker films show a behavior very close to that of
bulk~\cite{Biswas01} whereas substrate induced effects become more
important in thinner films.~\cite{Walter00}

The structural differences between La$_{2/3}$Ca$_{1/3}$MnO$_{3}$,
which is orthorhombically distorted, and
La$_{2/3}$Sr$_{1/3}$MnO$_{3}$ with its  rhombohedral symmetry
represent promising conditions to distinguish between competing
polaronic models. We discuss those models which adequately reproduce
the low energy side of the resonance, as these excitations
characterize the nature of the observed polarons.

\section{Experimental Analysis}

\subsection{Details and Characterization of the Films}

The thin films were grown by a standard pulsed laser deposition
technique\cite{Christey94} on plane-parallel $10 \times 10$~mm$^2$
single crystalline substrates: LSMO on
(LaAlO$_{3}$)$_{0.3}$(Sr$_{2}$AlTaO$_{5}$)$_{0.7}$ (LSAT) and LCMO
on NdGaO$_{3}$. Thin films with thickness between 200~nm and 400~nm
were investigated. The nominale lattice mismatch is $<$0.1\% for
both.

\begin{figure}[t]
\centering\vspace{-10mm}
\includegraphics[width=.5\textwidth,clip,angle=0]{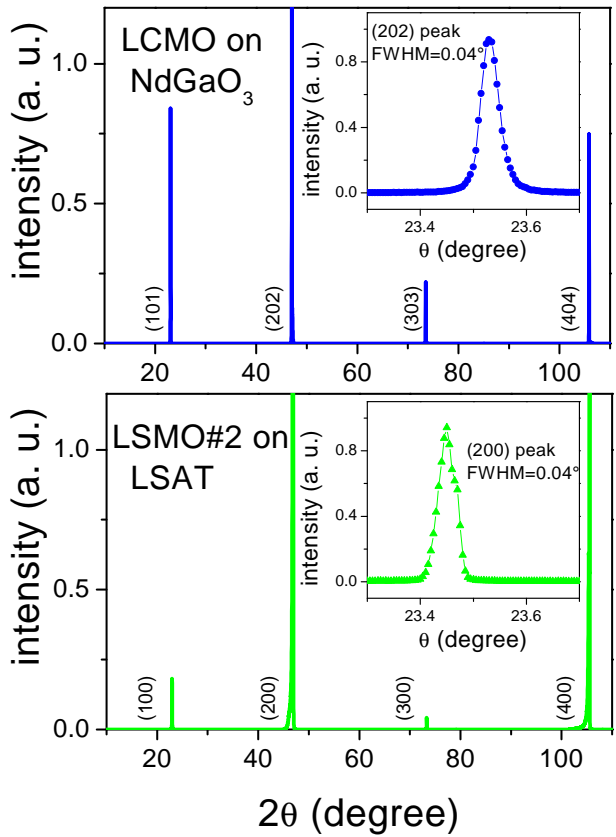}
\vspace{-5mm} \caption[]{\label{x-ray}  (Color online) Room
temperature $\Theta-2\Theta$ XRD pattern for LCMO on NdGaO$_{3}$
(upper panel) and for LSMO on LSAT (lower panel). The insets show
the rocking curve of the (202) peak and the (200) peak,
respectively.}
\end{figure}

X-ray diffraction (XRD) measurements were performed using a Bruker
D8 Discover four circle x-ray diffractometer. In Fig.~\ref{x-ray} we
show x-ray diffraction patterns ($\Theta-2\Theta$) obtained for LCMO
on NdGaO3 (upper panel) and LSMO on LSAT (lower panel). For
LCMO/NdGaO3 the XRD displays only (\textit{l{\rm 0}l}) diffraction
peaks for Pnma and for LSMO/LSAT only (\textit{l}00) diffraction
peaks for Pm3m symmetry. For both compounds there exist no
indications of any impurity phase. This points to the highly
oriented crystal growth of the films. The insets of Fig.~\ref{x-ray}
exhibit the rocking curves of the (202) peak for LCMO and the (200)
peak for LSMO. Both films show a very narrow full width at half
maximum (FWHM) of the rocking curve of 0.04$^\circ$ which give
evidence for a high quality crystalline structure of the films.

Reflectivity measurements of the film-substrate system were
carried out in the range of  50 to $4 \times 10^{4}$~cm$^{-1}$
using the Fourier transform spectrometers Bruker IFS 113v and
Bruker IFS 66v/S. Below room temperature the measurements were
performed with a He-cryostat where the sample is placed in a
static He exchange gas. The spectra at higher temperatures were
taken in a home made oven in which the sample was exposed to a
continuous flow of heated nitrogen gas. In addition, a
Mach-Zehnder interferometer was used for ac measurements between
10 and 30~cm$^{-1}$ which allowed to obtain both the intensity and
the phase shift of the wave transmitted through the films on the
substrate. Such transmission measurements of thin films have
proven to be a powerful method for studying the electrodynamic
properties of highly conductive thin films at low
frequencies.\cite{Kozlov98} The combined data sets improve the
Kramers-Kronig (KK) analysis of the film-substrate system in the
complete frequency range. For the analysis, the reflectivity
$R(\omega)$ was extrapolated by a constant for frequencies  below
10~cm$^{-1}$. For the high frequency region, the reflectivity
between $4 \times 10^{4}$~cm$^{-1}$ and 10$^{6}$~cm$^{-1}$ was
extrapolated by a $\omega^{-1.5}$ law. Above, a $\omega^{-4}$
dependence was assumed. The same procedure was utilized in
separate experiments for the respective substrate. Applying the
Fresnel optical coefficients for a two-layer system, the optical
conductivity of the film was evaluated. A more detailed
description of the data analysis is given in
Ref.~\onlinecite{HeavensBook}.

\begin{figure}[t]
\centering\vspace{0mm}
\includegraphics[width=.53\textwidth,clip,angle=0]{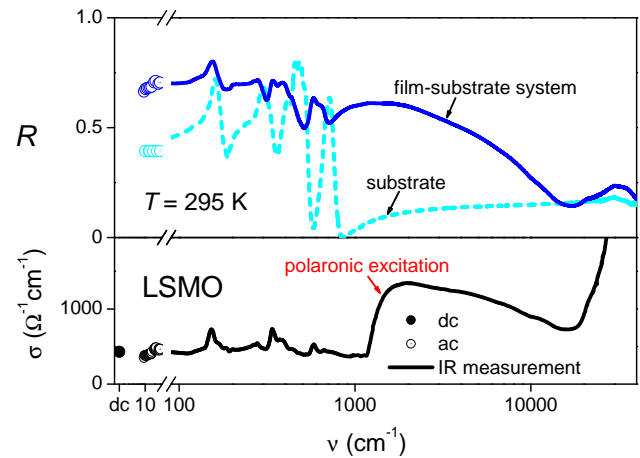}
\vspace{0mm} \caption[]{\label{film-substrat-vgl}  (Color online)
Upper panel: Reflectivity of the film-substrate system (LSMO on
LSAT) and of the substrate itself (LSAT) at 295~K. Lower panel:
The optical conductivity $\sigma$ of the film, evaluated from the
reflectivity data.}
\end{figure}

For room temperature the combined data sets of IR and ac
reflectivity $R$ are shown for the film-substrate system and for
the bare substrate in the upper panel of
Fig.~\ref{film-substrat-vgl} while the calculated conductivity
$\sigma$ is displayed in the lower panel. Most phonon resonances
of the substrate are located in the minima of the film-substrate
spectra. In the metallic phase the screening shields the lattice
modes of the substrate, therefore only resonances of the thin film
are visible. The temperature dependence of the phonon resonances
provides for an exact determination of the film properties
independently of the influence of the substrate. Moreover, the
resonances known from neutron scattering
measurements\cite{Reichardt99} are in good agreement with our
results.

The electrical dc-resistivity of the films was measured with a
standard four-point technique.\cite{Hartinger04c2} We estimate the
defect level in relation to the single crystal. Due to stronger
strain effects and a higher density of grain boundaries, the
defect level in LSMO~\#2 is about 12 times higher and in LSMO~\#1
about 24 times than in the single crystal. For LCMO a single
crystal was not available but the residual resistant is almost the
same as for LSMO~\#2.

The magnetization was obtained with a commercial Quantum Design
SQUID magnetometer. The measurements were carried out for the
substrate and for the deposited film. The data were reported in a
previous publication,\cite{Hartinger04c2} and a single crystal
measurement is found in Ref.~\onlinecite{Hartinger04c1} for
comparison. To characterize the ferromagnetic to paramagnetic
(FM-PM) and the metal-insulator (MI) transition of the samples, we
define $T_{C}$ as the magnetic Curie temperature and
$T_{\textrm{MI}}$ is determined from the maximum of the dc
resistivity curves. For LSMO the onset of spontaneous
magnetization appears at 345~K, well below $T_{\textrm{MI}}\approx
390$~K, while for LCMO $T_{C}\approx243$~K coincides with
$T_{\textrm{MI}}$.

\subsection{Experimental Results}

\begin{figure}[t]
\centering\vspace{0mm}
\includegraphics[width=.50\textwidth]{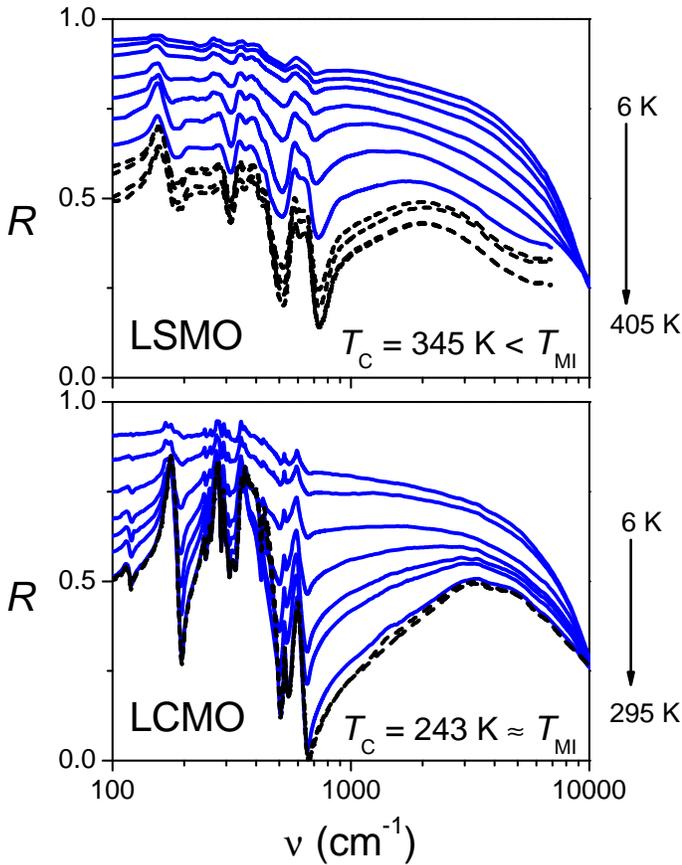}
\vspace{0mm} \caption[]{\label{R}(Color online) Reflectivity
spectra for LSMO on LSAT (upper panel) and LCMO on NdGaO$_{3}$
(lower panel) over a temperature range from the metallic to the
insulating phase. The solid lines represent spectra in the FM
phase, the broken lines in the PM phase.}
\end{figure}

The reflectivity spectra for frequencies below the optical
transitions are displayed in Fig.~\ref{R} for LSMO (upper
panel)~\cite{comment0} and for LCMO (lower panel).
The solid lines represent the spectra for temperatures in the
metallic FM pase while the broken lines refer to those in the PM
phase. The series of sharp peaks below $\sim 700$~cm$^{-1}$ were
identified as phonon modes. A detailed analysis of the
phonon excitations in these films has been presented in a previous
article.\cite{Hartinger04c2} The identification of the infrared-active
phonons attests the quality of the thin films. In the following
we focus on the excitation above $\sim 700$~cm$^{-1}$, which has
been assigned to polarons
before.\cite{Kaplan96,Okimoto97,Jung98,Kim98,Saitoh99,Lee99,Quijada98,Machida98,Hartinger04}

\begin{figure}[t]
\vspace{0mm} \centering
\includegraphics[width=.48\textwidth,clip]{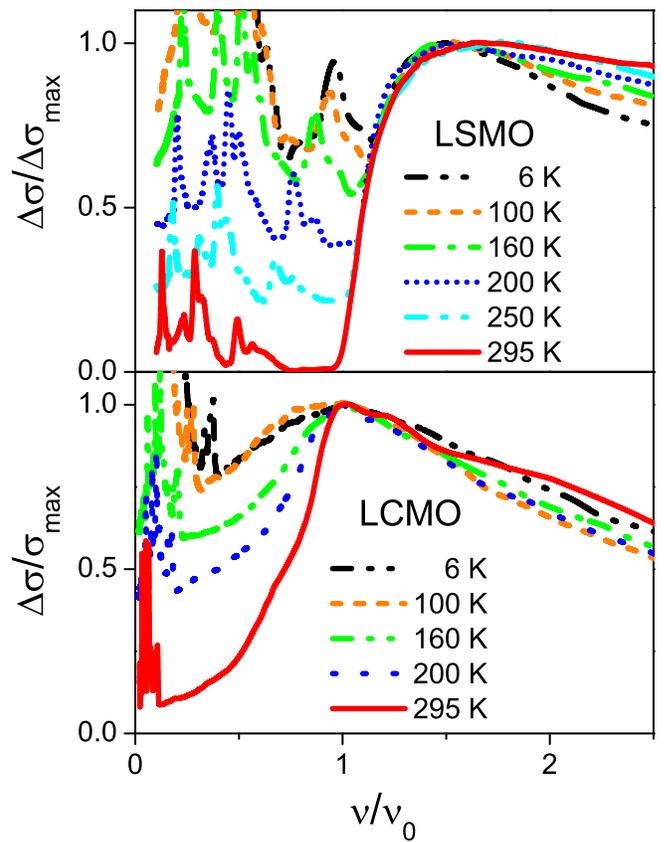}
\vspace{0mm} \caption[]{\label{scaling} (Color online) Rescaled
optical conductivity of LSMO (upper panel) and LCMO (lower panel):
the optical conductivity is scaled by the maximum of the MIR
resonance. A constant background of $\sigma_{\rm
bg}=$370~$\Omega^{-1}$cm$^{-1}$ for $\Delta\sigma$ is subtracted
for LSMO (for details, see main text). The frequency is scaled by
the threshold value $\nu_0$ of the onset of the polaronic peak for
LSMO, while for LCMO by the peak maximum. Clearly, the scaling
behavior of the low frequency slope is missing for the LCMO
spectra.} \vspace{0mm}
\end{figure}

For increasing temperatures the polaronic resonance in the optical
conductivity continuously shifts to higher frequencies and,
moreover, a loss of spectral weight is observed
--- as has been discussed in Ref.~\onlinecite{Hartinger04}.
Both quantities, frequency and amplitude of the resonance, will be listed
in Tabs.~\ref{bindungsenergie} and \ref{TKFit} for LCMO and LSMO,
respectively, linked to specific theoretical models.
There is a remarkable difference in the shape of the polaron peak
for LSMO and LCMO. In LSMO the shape is asymmetric and displays a steep slope below the
peak position, whereas for LCMO the broader hump is more symmetric
about its maximum and a low frequency threshold is entirely missing.

When we scale $\sigma(\nu)$ by a typical energy $\nu_0$ of the
polaronic resonance we find a universal low-energy slope for the
LSMO spectra (upper panel of Fig.~\ref{scaling}) whereas this
scaling behavior is absent in the LCMO spectra (Fig.~\ref{scaling},
lower panel). For this presentation, $\nu_0$ is the threshold on the
low-energy side of the MIR-resonance of the LSMO spectra and the
frequency at the  maximum of the MIR resonance of the LCMO spectra,
respectively. To retrieve the threshold frequency $\nu_0$ for LSMO,
one has to take into account that a background adds to the polaronic
absorption close to the threshold. We assume that the 295~K data
approximately display the background for the considered frequency
window: the phonon resonances are sufficiently far below $\nu_0$ so
that they should interfere very little with the polaronic resonance,
and the frequency dependence of the intensity of excitations
responsible for the background is negligible for the considered
frequency window in which we address the scaling behavior.  We then
extrapolate the low frequency slope of the curves at all other
temperatures to this ``zero line'' in order to identify  $\nu_0$.
Strictly speaking, $\nu_0$ depends slightly on the extra\-polation
procedure. This qualitative distinction of the LSMO and LCMO spectra
has to be addressed in a thorough theoretical modelling.

\section{Further evidence for small and large polaron scenarios}

\subsection{Holstein SP hopping}

The MIR optical conductivity displays distinct polaronic excitations
for LSMO and LCMO. It is expected that this distinction
also manifests itself in a disparate dc-resistivity for the two types
of manganite films.

\begin{figure}[b]
\centering
\includegraphics[width=.5\textwidth,clip,angle=0]{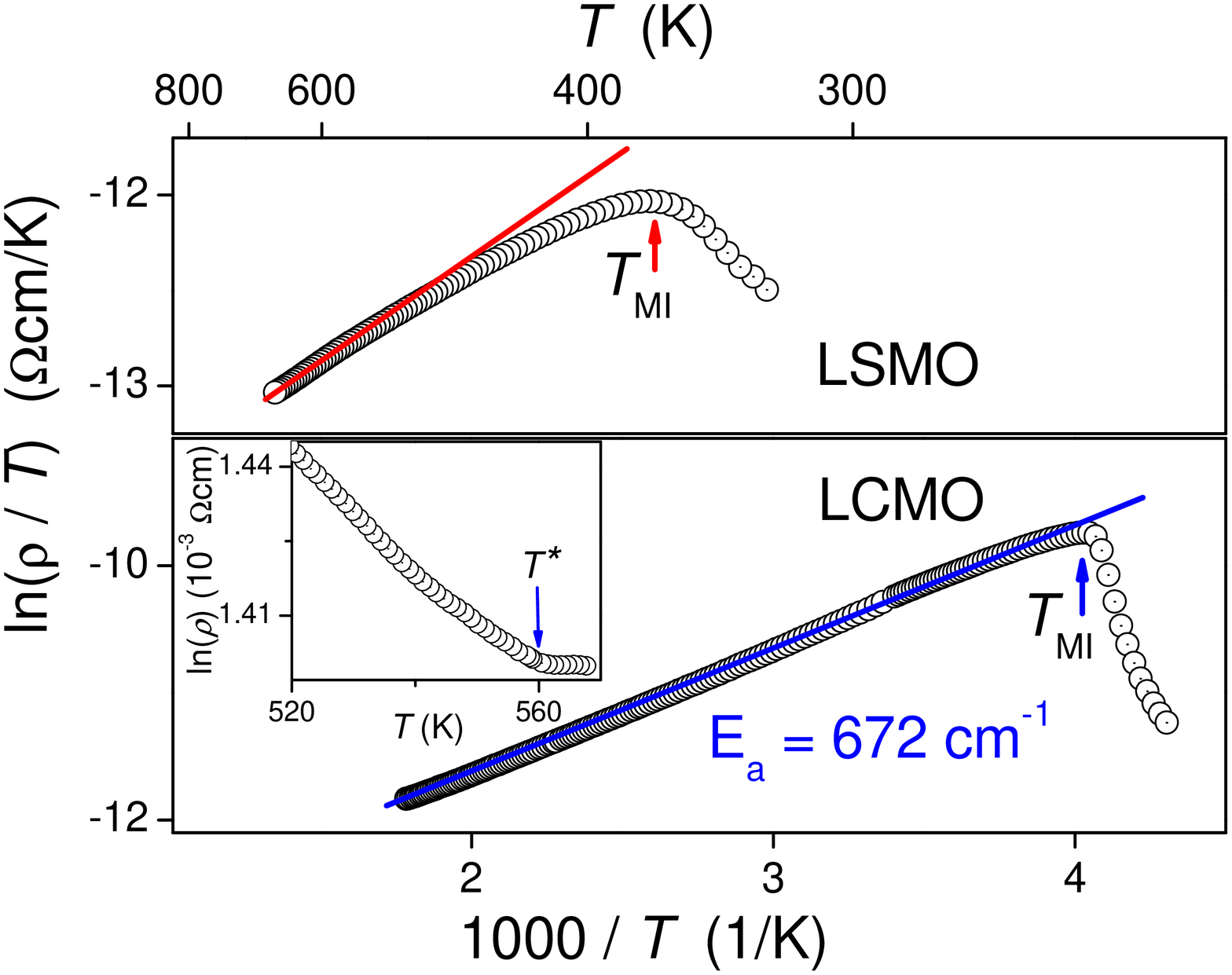}
\vspace{0mm} \caption[]{\label{invrho}(Color online) High
temperature dc-resistivity of LSMO (upper panel) and LCMO (lower
panel). The solid lines are the fit of the adiabatic SP model. For
LCMO the curve is linear and corresponds to polaron conduction with
an activation energy $E_{a} = 672$~cm$^{-1}$. The inset displays a
kink in the dc-resistivity of LCMO at 560~K which we identify with
$T^{*}$ (see main text).}
\end{figure}

In Fig.~\ref{invrho} we show the high temperature dc-resistivity of
LSMO (upper panel) and LCMO (lower panel) in a $\ln (\rho /T)$ vs.\
$1/T$ scale. In the lower plot a linear dependence for the latter
compound above $T_{\rm MI}$ is clearly visible. Slightly above
$T_{\rm MI}$, which coincides with the magnetic ordering
temperature, deviations from the thermally activated behavior occur.
It is widely accepted that the electron-phonon coupling is
sufficiently large to localize the carriers in the paramagnetic
phase --- presumably in combination with other localization
mechanisms. Over a temperature range from $255-560$~K the electrical
dc-resistivity is well described by the adiabatic SP
model~\cite{Emin69} (lower panel) with a dc activation energy $E_{a}
= 672$~cm$^{-1}$, which is in agreement with previously reported
results.\cite{Jaime96,Snyder96,Worledge96,Zhao00} In general the
binding energy $E_{b}$ of the SP in the optical data (see
Tab.~\ref{bindungsenergie}) is related to the activation energy of
the transport data by $E_{b}=2E_{a}$. The maximum in the optical
conductivity in the paramagnetic phase ($T=250$~K) is located around
$6040$~cm$^{-1}$, which is rather $8E_{a}$ than the theoretically
expected $2E_{b}=4E_{a}$. The discrepancy between the transport
value and the optical value is not surprising since it was also
observed in other polaronic systems\cite{Bi93} (including the
classical small polaron material TiO${}_2$; the difference has been
discussed by Reik\cite{Reik72} and Austin,\cite{Austin69} and for
the manganites by Hohenadler and Edwards\cite{Hohenadler02}).

Neutron scattering measurements for single crystal sample were
interpreted in terms of ''single polarons'' and ''correlated
polarons'' (see, e.g.\ Ref.~\onlinecite{Dai00,Nelson01}) --- the
existence of which is supposed to be consistent with a phase
separation scenario:\cite{Dagotto05} Alvarez \textit{et
al.}~\cite{Alvarez02} predict uncorrelated magnetic clusters in a
temperature range $T_{C} \leq T \leq T^{*}$. Below $T_{C}$ the
clusters align their moments. Above $T^{*}$ single polarons are
expected to dominate the transport.\cite{Dagotto05} In fact, for
LCMO we find a kink in the high temperature dc-resistivity data (see
$\ln \rho(T)$ in the inset of Fig.~\ref{invrho}). We tentatively
identify this kink at 560~K with $T^{*}$. This observation is
qualitatively consistent with film and single crystal results
reported in Ref.~\onlinecite{Snyder96} but we question their
interpretation according to which the kink is a signature of a
structural transition. Rather it appears that the different slopes
depend on various scattering processes of correlated and single
polarons, and the temperature $T^{*}$ characterizes the crossover.

Previously, it has been reported that correlated polarons have been
found only in the orthorhombic phase (LCMO).\cite{Kiryukhin03} This
observation may be related to the distinct high temperature behavior
in LCMO and LSMO. For LSMO we identified LP-type excitations from the optical
conductivity data. Figure~\ref{invrho} (upper panel) shows the high
temperature dc-resistivity from $330 - 750$~K. Clearly, the
adiabatic SP model (solid line) provides a poor fit to the
experimental data (compare the discussion in Ref.~\onlinecite{Geller01}).

The temperature dependence of the JT distortion for
La$_{0.7}$Sr$_{0.3}$MnO$_{3}$ has been investigated more recently by
Mannella \textit{et al.}\cite{Mannella04,Mannella05} with x-ray
absorption spectroscopy.  They provide direct experimental evidence
for the presence of polarons in the paramagnetic state and analyze
the dopant dependence of the JT distortion. Specifically, the variance
of the Mn-O bond length distribution in the Sr doped sample was shown
to be only half that in the Ca doped sample.

\subsection{Charge Carrier Mobility}

\begin{figure}[b]
\centering
\includegraphics[width=.52\textwidth,clip]{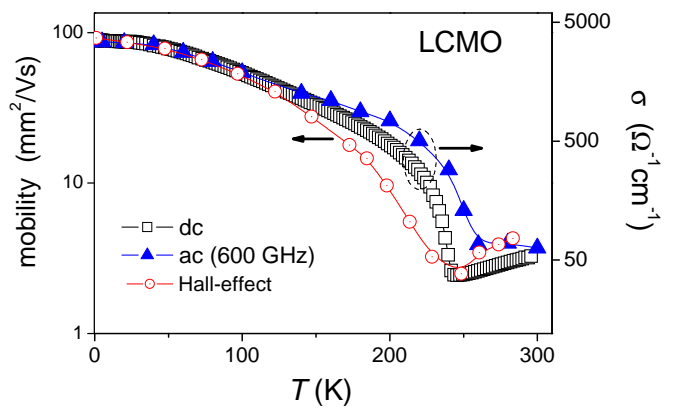}
\vspace{0mm} \caption[]{\label{ca-dc-ac-mu}(Color online) For LCMO
the temperature behavior of dc and ac electrical conductivity $\sigma$
(right axis) corresponds to the temperature dependence of the
mobility (left axis) obtained from Hall-effect
measurements.\cite{Jakob98}
The similarity of the data indicates a common physical origin.}
\end{figure}

Further support for the SP concept in LCMO and the LP ansatz in LSMO
comes from the literature on the polaron mobility.
Figure~\ref{ca-dc-ac-mu} compares the temperature dependence of the
dc- and ac-conductivities (right axis) with the mobility (left axis)
obtained by Hall-effect measurements by Jakob \textit{et
al.}\cite{Jakob98} The carrier mobility decreases with increasing
temperature in the FM phase. Above $T_{\rm MI}$ the mobility
increases slightly. All three measurements show the same overall
temperature behavior. The deviation close to $T_{\rm MI}$ may be due
to different coupling mechanisms of the local transport properties
to critical magnetic fluctuations.

The decrease of the mobility in the FM phase corresponds to the
increase of the effective electron-phonon coupling (polaron energy
over bandwidth). Moreover, the spectral weight of the polaronic
excitation is reduced on increasing temperature.\cite{Hartinger04}
As a consequence, with decreasing lattice distortions, the tendency
to charge localization is reduced.

For LSMO only few Hall effect studies have been
reported, none of them providing values of the polaron mobility
$\mu_{H}$.\cite{Geller01,Mandal98,Asamitsu98} Therefore, only a
rough estimate as $R_{H}/\rho_{xx} \approx 400$~mm$^{2}$/Vs at 4.2~K
is possible,\cite{Asamitsu98} which is four times larger than in
LCMO (Fig.~\ref{ca-dc-ac-mu}). The ratio of the mobilities
corresponds well to the ratio of the effective masses of
specific-heat measurements,\cite{Wang01} and reveals the expected
tendency. We propose that optical and Hall transport consistently
support the applicability of a LP-type approach. This interpretation
is also in agreement with neutron scattering results where the
spacial extension of the polaron covers several lattice constants at
low temperatures: Louca and Egami \cite{Louca99}  investigated
deviations of the local structure from the average crystal structure
in La$_{1-x}$Sr$_{x}$MnO$_{3}$ by pulsed neutron scattering with
pair distribution function analysis. X-ray absorption fine structure
measurements\cite{Shibata03,Mannella04} differ for the metallic
phase quantitatively, but all groups identify a
temperature-dependent lattice distortion.

\section{Discussion of the MIR spectra}

Early on the double exchange (DE),\cite{Zener51,Anderson55} in
combination with a strong Hund coupling, was identified as a key
element to explain the interrelation between metallicity and
ferromagnetism in the doped manganites. It has only recently been
understood that further basic mechanisms have to be an integral
part of a comprehensive model that can explain magnetotransport
properties and the transition to the paramagnetic phase. Millis,
Littlewood and Shraiman\cite{Millis95} elaborated that the DE
mechanism is not sufficient to reproduce the measured resistivity
for temperatures of the order of \TC. They suggested that, in the
presence of a strong electron-phonon coupling, polaronic effects
have to be included. Dagotto, Hotta and Moreo\cite{Dagotto01}
promoted that intrinsic inhomogeneities are generated through the
cooperative effect of phase separation tendencies, long-range
Coulomb interaction and disorder, and these inhomogeneities are
supposed to dominate the transport behavior.

In the following discussion on the nature of the MIR resonance, we
will consider only the polaronic models with no refinement of
electron-electron interactions or disorder physics. This disregard
of some of the essential ingredients of a comprehensive
microscopic model for manganites is in general not justified.
However, a solution to such a ``comprehensive model'' is at
present not available, certainly not for an authoritative
evaluation of the MIR optical conductivity. On the other hand,
some of the electronic correlations missing in the polaron models
would enter only as effective parameters, such as the
temperature-dependent kinetic energy, or the projection on one spin
direction. Consequently, these correlations do not necessarily
have to be disentangled from a more microscopic modelling.
Finally, the purpose of our discussion is also a critical review
of the polaronic models which have been previously related to
the resonance. We reemphasize that in the considered thin
films, the resonance is more pronounced and the shape is well
resolved so that a refined analysis is now attainable with our
data.

\begin{figure}[t]
\vspace{0mm} \centering
\includegraphics[width=.50\textwidth,clip]{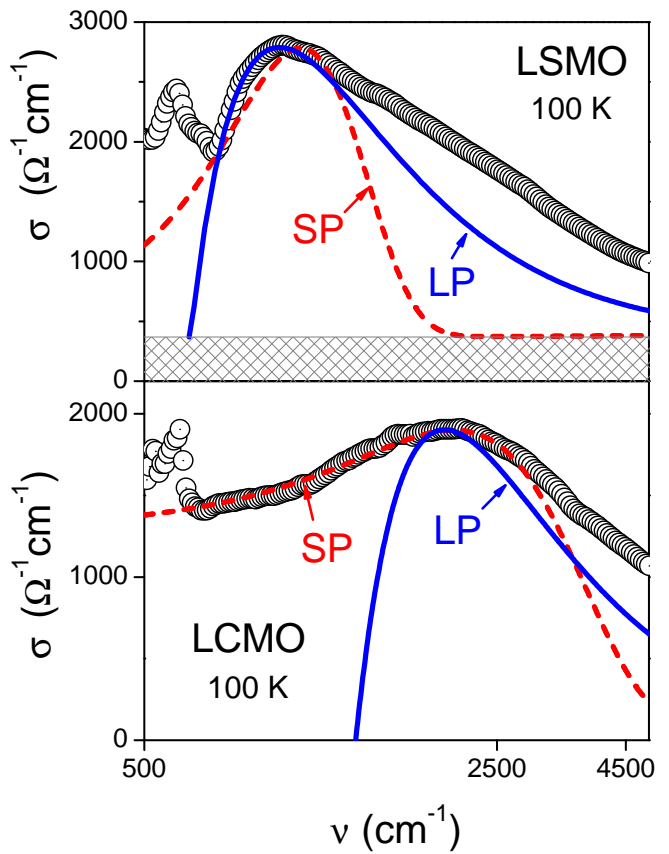}
\vspace{0mm} \caption[]{\label{fit}(Color online) The optical
conductivity spectra of LSMO (upper panel) and LCMO (lower panel) at
100~K (data points) on a logarithmic frequency scale. For a
preliminary assignment to large polarons (LP, solid line) and small
polarons (SP, dashed line), Eqs.~(\ref{eq:ScalingForm}) and
(\ref{eq:polaron}) have been used for the respective fits.}
\end{figure}

A small polaron in the regime of sufficiently strong electron-lattice
coupling is excited from a site-localized state to another
well-localized state at an adjacent site.\cite{comment1} The
maximum of the absorption, that is the most probable process, is
the Franck-Condon like transition with frequency $\omega_m$ four
times the activation energy for the intersite
transition.\cite{Reik72,MahanBook} The MIR conductivity then has
the form of an asymmetric Gaussian peak centered around
$\omega_m$. Whereas the optical conductivity of LCMO approaches
this shape and therefore is identified as a small polaron, the
optical conductivity of LSMO does not (compare Fig.~\ref{fit}).

A large polaron is observed usually in the weak coupling regime.
Then, to first order in the electron-lattice coupling, the optical
absorption is the excitation of a phonon with a simultaneous
excitation of the charge carrier GLF\cite{Gurevich62}.
Consequently, one finds a threshold behavior at the (optical)
phonon frequency with a sharp increase in the absorption rate due
to the available scattering states and a slow decrease at the
high-energy side controlled by a $1/\omega^3$-factor, most easily
identified in the force-force correlation function form of the
conductivity.\cite{MahanBook} For intermediate to strong
electron-lattice coupling, if the polaron is in a self-trapped
state, the threshold is determined by the polaron binding energy,
with the charge carrier scattering into an extended state
(Emin\cite{Emin93} assumes a free-carrier state). The initial upturn of
the absorption spectrum is again given by the phase space of the
available scattering states. The shape of the observed optical
conductivity appears to be qualitatively similar to that in the
GLF- and Emin-approaches, and we tentatively identify the
polaronic resonance for LSMO as the excitation of a large polaron
(see Fig.~\ref{fit}). The validity of such models to
the manganite films will be investigated in the following
sections.

The asymmetric shape of the MIR resonance with the steep increase
on the low-energy side is quite spectacular for the investigated
LSMO films. As argued before, the upturn of the optical
conductivity above the threshold is controlled by phase space
considerations. For example, if the electronic states created by
the absorption process are particle-hole excitations close to a
Fermi edge, the number of such states increases linearly with the
energetic distance (frequency) to the edge. Whatever the details,
one may expect a scaling behavior of LSMO, at least for the
initial increase above the threshold. This scaling would signify
that the low energy scattering processes are of the same origin,
independent of temperature, related band narrowing and other
energy scales. In fact, a universal low-frequency slope has been verified for
the LSMO spectra (Fig.~\ref{scaling}, upper panel) but it is absent
for LCMO which neither displays a threshold on the low-energy side of
the MIR-resonance ( Fig.~\ref{scaling}, lower panel).

\subsection{Small polarons in LCMO}

Various experimental data sets on both, ferromagnetic and paramagnetic
manganites, have been interpreted in terms of the Holstein
small-polaron theory\cite{Emin69} (see our previous section, and also the critical
discussion on this issue in Ref.~\onlinecite{Edwards02}). An extension
of the standard SP optical conductivity
formula\cite{Reik72,Emin75} has been proposed for the evaluation of
the low-temperature optical spectroscopy:\cite{Puchkov95,Yoon98}

\begin{equation}
\label{eq:polaron} \sigma(\omega,T)= \sigma(0,T) \frac{\sinh (4
E_{b} \hbar\omega / \Delta^2)}{4 E_{b} \hbar\omega / \Delta^2} \,
e^{-(\hbar\omega)^{2}/\Delta^{2}}  \quad
\end{equation}
\begin{equation*}
\Delta\equiv 2 \sqrt{2 E_{b} E_{\rm vib}} \quad
\end{equation*}
Here $E_{b}$ is the SP binding energy,
$\sigma(0,T)$ is the dc conductivity, $\Delta$ signifies a
broadening, and $E_{\rm vib}$ is the characteristic vibrational
energy, which is $E_{\rm vib}=k_{B}T$ in the high-temperature
regime and $E_{\rm vib}=\hbar \omega_{\rm ph}/2$ at low
temperatures ($k_B T < \hbar\omega_{\rm ph}$).
Eq.~(\ref{eq:polaron}) constitutes a broad resonance with a
maximum at $\hbar\omega_m \approx 2E_{b}$. For high temperatures,
the dc-conductivity results from the thermal activation over the
potential barrier ($E_b/2$) between adjacent sites and the optical
conductivity peaks at the frequency ($2E_b$) of the
Franck-Condon-like transition for neighboring potential energy
curves, which are approximately parabolic in the configurational
coordinates.  For low temperatures, the processes are similar, however the phonons which
participate in the activated hopping are not thermal.

The dashed lines in Fig.~\ref{fit} present the optical
conductivity of a SP, Eq.~(\ref{eq:polaron}). The shape for low
frequencies, including the maximum, is well reproduced for LCMO.
The SP model nicely describes the experimental results for 100~K
using $2E_{b}=2820~$cm$^{-1}$ and a characteristic phonon energy
$E_{\rm vib}=270~$cm$^{-1}$. The disagreement between theory and
experiment for higher frequencies is not unexpected (it has also
been observed for the classical small polaron material TiO$_{2}$,
and several mechanisms were proposed to account for the
difference;\cite{Kudinov70} see also the SP evaluation for the
2\textit{d} nickelates and cuprates\cite{Bi93}).
As seen from Tab.~\ref{bindungsenergie} and Tab.~\ref{GaussFit},
with rising temperature the polaron formation energy increases.
For LCMO the binding energy ($2E_{b}$) rises from 2210~cm$^{-1}$
(6~K) to 6040~cm$^{-1}$ (250~K). Above the metal-to-insulator
transition at $T_{\rm MI}$, the polaron formation energy remains
constant.

\begin{table}[b]
\caption[]{Fit parameters of the optical conductivity for LCMO at
different temperature using the SP formula,
Eq.~\ref{eq:polaron}.} \label{bindungsenergie} \vspace{.2cm}
\centering
\begin{ruledtabular}
\begin{tabular}{lccc cc cc cc cc cc cc}
  $T$ (K) && 6 && 100 && 160 && 200 && 220 && 240 && 250\\
  \hline
  $\sigma_{0}$ ($\Omega^{-1}$cm$^{-1}$)&& 1410 && 1270 && 930 && 685 && 410 && 130 && 90\\
  $2E_{\rm b}$ (cm$^{-1}$)&& 2210 && 2820 && 3720 && 4080 && 5020 && 5700 && 6040\\
  $E_{\rm vib}$ (cm$^{-1}$)&& 217 && 270 && 340 && 335 && 330 && 270 && 270\\
\end{tabular}
\end{ruledtabular}
\end{table}

Whereas one may expect that the {\it low-energy} SP absorption is
well explained by the processes described above --- since they are
local ---, one has to consider the consequences of a high polaron
concentration and Fermi energy for the {\it high-energy} SP
absorption in order to understand the slow decay observed in
experiments. Especially for temperatures of the order of $T_{C}$
polaron-polaron interaction appears to be
strong,\cite{Dai00,Nelson01} however a relation to the high-energy
slope of the polaronic MIR resonance is not evident.

An explanation of the high energy decay and of the temperature
dependence of the polaron binding energy is beyond the scope of this
paper. A transfer of spectral weight towards higher energies was
discussed within a dynamical mean field theory of the Holstein model
by Fratini {\it et al.},\cite{Fratini01} however, their treatment
only applies to a dilute nondegenerate polaron gas. The shape and
position of the polaronic MIR resonance for a model with Fermi edge
and a finite concentration of polarons has been investigated by
Millis, Mueller and Shraiman (MMS).\cite{Millis96} They study the DE
model with a JT form for the electron-phonon coupling. The local
phonons are taken as classical harmonic oscillators, and the
electronic correlations are evaluated within the DMFT scheme. For
half filling ($n_{e_g}=1$, with the $e_g$ orbitals singly occupied)
and intermediate coupling, MMS compare their optical conductivity
favorably with the data of Okimoto {\it et
al.}\cite{Okimoto95,Okimoto97} for
{La$_{0.875}$Sr$_{.175}$MnO$_{3}$} --- and for somewhat stronger
coupling with the data of Kaplan {\it et al.}\cite{Kaplan96} for
{Nd$_{0.7}$Sr$_{.3}$MnO$_{3}$}. Also our optical data agree
qualitatively with the latter results of MMS though the high
frequency side again decays too fast in the model evaluation. The
broad peak is due to transitions between the two JT split levels
with a strongly fluctuating phonon coordinate. The actual results
for the evaluation of doped systems are less favorable: a second
peak below is formed which is due to transitions to unoccupied sites
with no JT splitting. It appears to contradict the observations.
However, MMS argue that further electronic correlations which have
not been included can remedy this shortcoming.

Some of the quantum fluctuations missing in MMS are included in the
``many-body CPA'' treatment of the Holstein-DE model\cite{commentCPA} for which
Hohenadler and Edwards\cite{Hohenadler02} compare the MIR optical
conductivity with the measurements\cite{Kaplan96} for
{Nd$_{0.7}$Sr$_{.3}$MnO$_{3}$}. The shift of spectral weight to
lower frequencies with decreasing temperature is still insufficient.
Perroni {\it et al.}\cite{Perroni01} propose that their variational
Lang-Firsov treatment recovers the low temperature
behavior,\cite{comment7,comment8} including a Drude peak.

\subsection{Bipolaron theory for the optical conductivity}

The crossover from the paramagnetic to the ferromagnetic phase in
the optical absorption has been studied by Alexandrov and
Bratkovsky\cite{Alexandrov99} in terms of a bipolaron model in
which the optical intraband conductivity should be provided by two
different polaronic contributions originating from SP and
bipolarons with different binding energies. Accordingly, in the PM
phase the optical conductivity would be dominated by small
bipolaronic charge carriers, which break up at the magnetic
transition, leaving only SP which then dominate the MIR
excitations in the FM regime. As a result, a sudden transfer of
spectral weight at $T_{C}$ is predicted along with a sharp drop in
the dc-resistivity due to the charge carrier collapse. As such a
transfer is not confirmed by our data, we rather discard such a
scenario for LSMO and LCMO, even though the oxygen-isotope effect on
the thermoelectrical power in several ferromagnetic manganites
seems to provide experimental clues for bipolaron
formation.\cite{Zhao00}

\subsection{Large polarons in LSMO}
\label{sec:LP_in_LSMO}

We will now
inquire if models, which have been recently proposed in this context,
indeed reproduce the observed scaling and threshold behavior for
LSMO. For this purpose, we organize this Section~\ref{sec:LP_in_LSMO}
as follows:

In a first approach we will use a well-controlled weak coupling scheme
(see Sec.~\ref{sec:WeakCoupling}) which follows from a RPA-evaluation
of the Hubbard model to which a Fr\"ohlich type electron-phonon
coupling has been added. As we only consider the ferromagnetic
metallic state we restrict the evaluation to the spin polarized sector
(spinless fermions). The local interaction $U$ signifies the on-site
interaction between charge carriers in the two $e_g$-orbitals.
The RPA-results well approximate the low-frequency slope of the polaronic
resonance and reproduce the observed scaling. In fact, the model replicates
the most prominent and important characteristics of the polaronic spectra in LSMO.
However it fails to treat the high energy features adequately.

In a second approach we investigate intermediate and strong
coupling schemes for large polarons (see the end of
Sec.~\ref{sec:WeakCoupling} and the Section~\ref{sec:SelfTrapped}
on {\it self-trapped polarons}). The latter scheme was proposed
recently,\cite{Kim98} in order to identify the MIR absorption
processes in bulk manganites as excitations from a localized
self-trapped state to free-carrier continuum states (the authors
of Ref.~\onlinecite{Kim98} refer to the model presented by Emin in
Ref.~\onlinecite{Emin93}). The shape of the polaronic resonance in
LSMO is well reproduced for the low energy side and the model may
account for a shift of the threshold energy with increased binding
(at higher temperatures). However the theoretical (strong
coupling) estimate of the polaron radius differs significantly
from the radius which fits the experimental data. Moreover, the
approach does not apply when a Fermi surface of the (free) charge
carriers is formed.

Finally, in Sec.~\ref{sec:FermiEdge} we discuss the consequences
when charge carriers are excited from a potential well to a Fermi
surface. The collective behavior associated with the response of
the Fermi sea to the transition determines the shape of the
resonance in the adiabatic limit (in which the potential well
decays slowly with respect to the fast electronic transition). A
full evaluation of this fascinating scenario is not yet available.

\subsubsection{Weak coupling scenario}
\label{sec:WeakCoupling}

\begin{figure}[t]
\vspace{0mm} \centering
\includegraphics[width=.48\textwidth,clip]{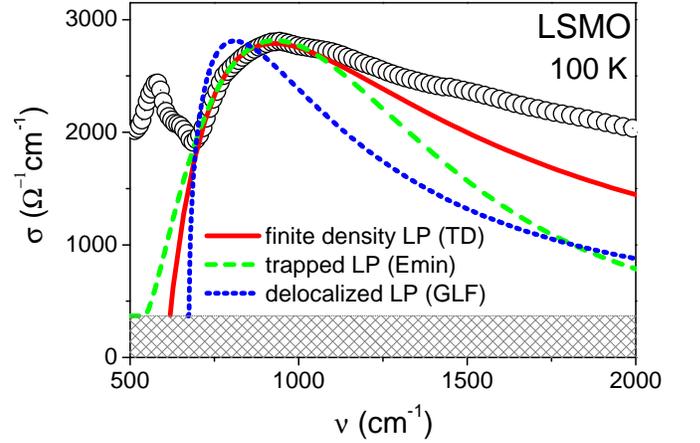}
\vspace{0mm} \caption[]{\label{theory} (Color online) Comparison
of the low temperature (100~K) optical conductivity (in the MIR
range of the polaronic excitations) to the results from various
model calculations: the continuous line refers to the weak
coupling approach of Tempere and Devreese modified for an on-site
Hubbard interaction, Eq.~(\ref{eq:HubbardSigma}); the dashed line
is the result of the phenomenological approach for self-trapped
large polarons by Emin, modified to Gaussian localized states,
Eq.~(\ref{eq:GaussianPolaron}); the dotted curve is the weak
coupling single-polaron result by Gurevich, Lang and Firsov,
Eq.~(\ref{eq:GLF}). The appropriate parameters for each of these
model evaluations are given in the main text.}
\end{figure}

The shape of the MIR resonance in the LSMO films is reminiscent of
the weak coupling result by GLF for Fr\"ohlich polarons (see, for
example, Ref.~\onlinecite{MahanBook} and  Eq.~(\ref{eq:GLF})
below). A closer inspection reveals that the width of the GLF
absorption is too small and the low-frequency slope is too steep
to fit the LSMO data (see Fig.~\ref{theory}). This failure is not
unexpected since we have $10^{21}$--$10^{22}$ charge carriers per
cm${}^3$, as identified from Hall measurements.\cite{Mandal98} An
extension of the GLF weak-coupling polaron theory to a finite
concentration has been carried out by Tempere and
Devreese.\cite{Tempere01} In their work on {\it optical absorption
of an interacting many-polaron gas}, they found a remarkable
agreement between their theoretical prediction and the
experimental curve for Nd$_{2}$CuO$_{4-y}$\cite{Calvani01} in the
optical conductivity around 1000 cm$^{-1}$. We now transfer their
approach to the manganites.

The real part of the optical conductivity $\sigma(\omega)$ is
expressed through a momentum integral
over the dynamical structure factor $S(q,\omega)$ which is in three space
dimensions:\cite{MahanBook,Tempere01}
\begin{equation}
\label{eq:DevreeseSigma}
\sigma(\omega)= \alpha\, n_p \frac{2}{3} \frac{e^2}{m^2}
  \frac{(\hbar\omega_0)^2}{\pi\hbar\omega^3}
   \sqrt{\frac{\hbar}{2m\omega_0}}\int^\infty_0 dq q^2
   S(q,\omega-\omega_0)
\end{equation}
Here, $n_p$ is the polaron density, $m$ is the effective mass of
the charge carriers, $\alpha$ is the electron-lattice coupling and
$\hbar \omega_0=h \nu_0$ is the threshold energy for polaronic
absorption (in weak coupling, it is the energy of the respective
optical phonon). The dynamical structure factor is expressed in
terms of the dielectric constant
$\epsilon(q,\omega)$:\cite{MahanBook}

\begin{equation}
\label{eq:StructureFactor}
S(q,\omega)= \frac{\hbar}{n} \frac{q^2}{4\pi e^2}
   \; \Im\lbrack -\frac{1}{\epsilon(q,\omega)} \rbrack \, ,
\end{equation}
where $n$ is the density of charge carriers which should be identified
with the polaron density $n_p$ for the considered finite concentration. TD\ evaluated
Eq.~(\ref{eq:DevreeseSigma}) to linear order in $\alpha$ which implies
that $S(q,\omega)$ is taken to zeroth order in the electron-lattice
coupling. The dielectric constant is then approximated by  the
Lindhard (or RPA) expression for the homogeneous electron gas.
In this linear approximation (with respect to $\alpha$), TD argue
that, for the low-density limit, the GLF result is recovered. Indeed,
for fermions with one flavor, the structure factor approaches\cite{comment2}
$S(q,\omega)_{k_f\rightarrow 0} \rightarrow
2\pi\delta(\omega-\varepsilon_q/\hbar)$.
The optical conductivity assumes the limiting form of
GLF\cite{Gurevich62} (cf.~Fig~\ref{theory}):
\begin{equation}
\label{eq:GLF}
\sigma(\omega) _{k_f\rightarrow 0}\; \rightarrow\;
   \alpha\, n_p\, \frac{2}{3} \frac{e^2}{m}
       \frac{1}{\omega_0}
  \,(\frac{\omega_0}{\omega})^3
   \;\sqrt{\frac{\omega}{\omega_0} -1}
\end{equation}
For the manganites in the investigated ferromagnetic, metallic
regime with a carrier density of $10^{21}$--$10^{22}$~ cm${}^{-3}$,
we take the following standard simplifications for the
low-temperature evaluation: (i) only one spin direction prevails
(spinless fermions), due to DE and strong Hund coupling, (ii) there
are two degenerate orbital states (the two $e_g$-levels);
accordingly we deal with two flavors for the fermionic particles,
(iii) the interaction is represented by the local Hubbard $U$ for
two fermions on the same site in two $e_g$ orbital states, and not
by the long-range Coulomb interaction $v_q=4\pi e^2/q^2$, (iv) the
fermions acquire a quadratic dispersion. The last assumption may be
refined to the appropriate two-band dispersion of the $e_g$-levels.
However, for the discussion of the optical conductivity it is of
minor importance, especially when we primarily discuss the
low-frequency excitations with energy much less than the Fermi
energy. Furthermore, a dynamical JT effect is not implemented: it is
entirely missing in this weak coupling approach of TD.

The longitudinal dielectric response in the Hubbard model (with
quadratic dispersion) is in RPA:

\begin{eqnarray}
\label{eq:HubbardRPA} \epsilon(q,\omega ) = 1 &-&
\frac{3}{4}\frac{U}{E_f} \,
 \Biggl\{
   1-\frac{1}{(2x)^3}
       \Bigl\lbrack \nonumber\\
         & & \qquad\quad  (
            (x^2-y)^2 -4x^2
           ) \ln \frac{x^2 -y + 2 x}{x^2 -y -2 x} \nonumber\\
        & & \quad\quad + (
            (x^2+y)^2 -4x^2
           ) \ln \frac{x^2 +y + 2 x}{x^2 +y -2 x} \nonumber\\
       & & \qquad\qquad\qquad\quad \Bigl\rbrack
  \Biggl\}_{x\,\rightarrow\, q/k_f \atop
            y\,\rightarrow\, \hbar\omega/E_f + i0}
\end{eqnarray}

The related expression for the Coulomb gas\cite{Lindhard54} is
usually not presented in a closed form. Here care has been taken
to ensure that this closed expression for
the dielectric response is valid for all complex values of $y$.

The optical conductivity is (cf. the similar expression in
    Ref.~\onlinecite{Tempere01}  for the Coulomb gas model)
\begin{equation}
\label{eq:HubbardSigma}
\sigma(\omega)= \alpha\gamma\delta\int^\infty_0\!\!dx x^2\,
   \frac{4}{3 } \frac{E_f}{U} (\frac{\omega_0}{\omega})^3
         \;\Im\left[-\frac{1}{\epsilon(x,\omega\!-\!\omega_0)}\right] \, ,
\end{equation}
whereby
$$\gamma= n_p\, \frac{2}{3} \frac{e^2}{m} \frac{1}{\omega_0}$$
is the prefactor which already appears in the GLF-limit, and
$$\delta= \frac{3}{2\pi} k_f \sqrt{\frac{\hbar}{2m\omega_0}} $$
is a dimensionless prefactor which accounts for the finite charge
carrier density. The integration variable $x$ in
Eq.~(\ref{eq:HubbardSigma}) is the normalized wavevector
$x=q/k_f$. The dielectric response in the Hubbard model,
Eq.~(\ref{eq:HubbardRPA}), and the optical conductivity,
Eq.~(\ref{eq:HubbardSigma}), are in a simple way related to the
respective expressions for the Coulomb gas if we substitute  in
Eqs.~(\ref{eq:HubbardRPA}) and (\ref{eq:HubbardSigma}) the
interaction parameter by
\begin{equation}
\label{eq:HubbardCoulomb}
\frac{3}{4}\frac{U}{E_f}\;\longrightarrow\;\frac{\xi}{x^2} \, ,
\end{equation}
where $\xi$ is the dimensionless interaction strength of the
Coulomb gas, $\xi=e^2k_f/(\pi E_f)$.

Finally, we adjust the parameters tentatively in the following
way: a) the density of charge carriers is $6 \times
10^{21}$~cm$^{-3}$, which is the stoichiometric number of doped
holes and which is in the range of what has been estimated from
Hall measurements,\cite{Mandal98} b) we take $m$ three times the
bare electronic mass ($m=3 m_e$) which is consistent with specific
heat measurements,\cite{Salamon01} c) since $E_f=3270$~cm$^{-1}$
for the density of (a) and for the effective mass of (b), and
since we read from the 100~K data $\nu_0 = 617$~cm$^{-1}$ we fix
$\hbar \omega_0 /E_f = 0.19$, that is, we are in the adiabatic
regime, and d) $U/E_f=0.2$ in order to start, for a first
estimate, with small electronic interactions for which RPA is a
reasonable approach.

For this parameter set we find from Eq.~(\ref{eq:HubbardSigma}) a
shape of the polaronic resonance (see Fig.~\ref{theory}) which
reproduces well the experimental data for frequencies up to the
maximum. The high energy slope, however, decreases more rapidly
than in the measured optical conductivity.
The maximal optical conductivity is, in our
calculation, $\sigma_{\rm max}= \alpha \times 1.84 \times
10^3~\Omega^{-1}{\rm cm}^{-1}$ which should fit the data with
$\alpha\simeq 1.3$
--- whereby the shaded background in Fig.~\ref{theory} is to be subtracted.

If we had taken the Coulomb gas (as TD did for
Nd$_{2}$CuO$_{4-y}$) with the same density, effective mass and the
same ratio $\hbar \omega_0 /E_f$ as above for the Hubbard model,
we would find nearly the same line shape (not distinguishable in
the plot). The dimensionless interaction strength then is
$\xi=6.5$ and the plasma frequency is at $\hbar\omega_{\rm
plasma}/E_f=4.15$. Since the plasmon is consequently far outside
the frequency window of our investigation it is not surprising to
have similar results for the Coulomb-gas and Hubbard-model
approach for these reasonably large densities. However, the
maximal optical conductivity is $\sigma_{\rm max}= \alpha \times
0.11 \times 10^3~\Omega^{-1}{\rm cm}^{-1}$ in the Coulomb gas
case. This implies an electron-lattice coupling $\alpha$ of the
order of 20, which is probably too large\cite{comment3} by a
factor of 5 to 6 --- apart from the fact that the Coulomb gas
model is not quite appropriate for these manganites. We conclude
that the Hubbard model, rather than the Coulomb gas approach, is
the proper model Hamiltonian provided that also an electron-phonon
coupling term is added.

\begin{figure}[t]
\includegraphics[width=.5\textwidth,clip]{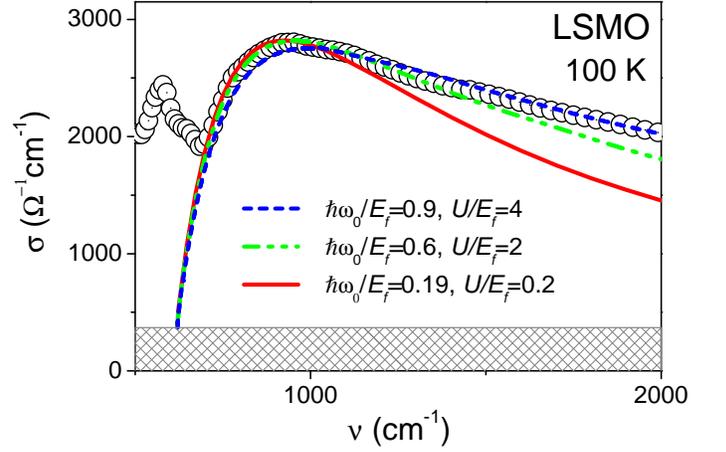}
\vspace{0mm} \caption[]{\label{theory-Hubbard}(Color online) In
the RPA-approach, Eq.~(\ref{eq:HubbardSigma}), the shape of the
optical conductivity depends on the model parameters $\hbar
\omega_0 /E_f$ and $U/E_f$. For the adiabatic weak-coupling case
($U/E_f=0.2$, $\hbar\omega_0/E_f=0.19$) the approximate form,
Eq.~(\ref{eq:ScalingForm}), is indistinguishable from the full
RPA-result, Eq.~(\ref{eq:HubbardSigma}). For details of the
comparison between the experimental 100~K data and the theoretical
results see the main text.}
\end{figure}

The solution, Eq.~(\ref{eq:HubbardSigma}), suggests to determine
the model parameters through an ``optimal fit'', now including the
high frequency slope. The shape of the optical conductivity in
this approach only depends on $\hbar\omega_0/E_f$, and the fine
tuning fixes the value of $U/E_f$. For small values of
$\hbar\omega_0/E_f$, such as for the above value 0.19, the
absorption shape is nearly independent of $U/E_f$, except that the
magnitude ($\sigma_{\rm max}$) scales down with increasing
$U/E_f$.\cite{comment3a} A better fit is found with larger values
of $\hbar\omega_0/E_f$. The optimal values are
$\hbar\omega_0/E_f=0.9$ and $U/E_f\simeq 4$ (see
Fig.~\ref{theory-Hubbard}). If we again assume that the density of
charge carriers is $6 \times 10^{21}$~cm$^{-3}$, we are
constrained to $m/m_e = 14.3$. Alternatively, we could fix $m/m_e
= 3$ but then we have to assume a strongly reduced polaron density
of $0.6 \times 10^{21}$~cm$^{-3}$. The drawback of this ``optimal
fit'' however is that $\sigma_{\rm max}/\alpha = 0.05 \times
10^3~\Omega^{-1}{\rm cm}^{-1}$, in the first case, and $0.02
\times 10^3~\Omega^{-1}{\rm cm}^{-1}$, in the second case, both of
them too small to be reasonably close to experiment. We deduce
from this investigation that the high energy side cannot be
satisfactorily captured within this model. The high energy slope
with its complex cooperative excitations has to be explored with
more involved theoretical concepts which necessarily depend on a larger
parameter set. However, this does not invalidate the evaluation
for the universal low energy side.

For the 6~K data, the low energy side of the polaronic resonance was fitted successfully in
our previous paper.\cite{Hartinger04,comment0} For the sample presented here
we find $\nu_{0}=593$~cm$^{-1}$, which implies that
$\hbar\omega_0/E_f=0.18$. With $U/E_f\simeq 0.2$ we have a nearly
identical fit to that of Fig.~4 in Ref.~\onlinecite{Hartinger04} which
generates sufficient spectral weight so that the electron-phonon
coupling $\alpha$ can be chosen to be of order 1. This situation is
qualitatively similar to that with the 100~K data. The same applies to
higher temperatures. We may assume that temperatures as high as 295~K
are still considerably smaller than the Fermi energy. Then temperature does not
explicitly enter the dynamical response but it is supposed to tune
the band width, and correspondingly the Fermi energy and effective
mass, and possibly $\omega_0$ (see the discussion below
on this last presumption).

{\it Assessment of the weak coupling approach.} From the above
discussion we deduce that a value of $\hbar \omega_0 /E_f \simeq
0.2$ and $U/E_f < 0.5$, is a  reasonable choice for low temperatures
to acquire sufficient optical weight in this approach. Thereby, the
precise values of $U/E_f$ and $\hbar \omega_0 /E_f$ just control
the overall magnitude of the  polaronic resonance provided
that $\hbar \omega_0 /E_f \ll 1$ applies.\cite{comment3b} Furthermore,
for the considered frequency range above the
threshold, we are always sufficiently close to the Fermi edge
in order to approximate the frequency dependence of the imaginary part of
$\epsilon(q,\omega-\omega_0)$ by $\omega-\omega_0$ which accounts for the number of
excited low-energy particle-hole pairs. Moreover, since
the frequency dependence of the absolute value of
$\epsilon(q,\omega-\omega_0)$ is weak, we expect the optical
conductivity, Eq.~(\ref{eq:HubbardSigma}), to approach
\begin{equation}
\label{eq:ScalingForm} \sigma(\omega)\;=\; A\,
\left(\frac{\omega_0}{\omega}\right)^3\;
(\frac{\omega}{\omega_0}-1) \, ,
\end{equation}
where $A$ is a frequency independent amplitude.
Indeed, this formula matches the frequency dependence of
$\sigma(\omega)$ so closely for the considered adiabatic regime of
$\hbar \omega_0 /E_f$ that it is indistinguishable
from the curve generated by the full weak coupling relation,
Eq.~\ref{eq:HubbardSigma} (Figs.~\ref{theory} and
\ref{theory-Hubbard}, continuous lines).

\begin{figure}[t]
\includegraphics[width=.5\textwidth,clip]{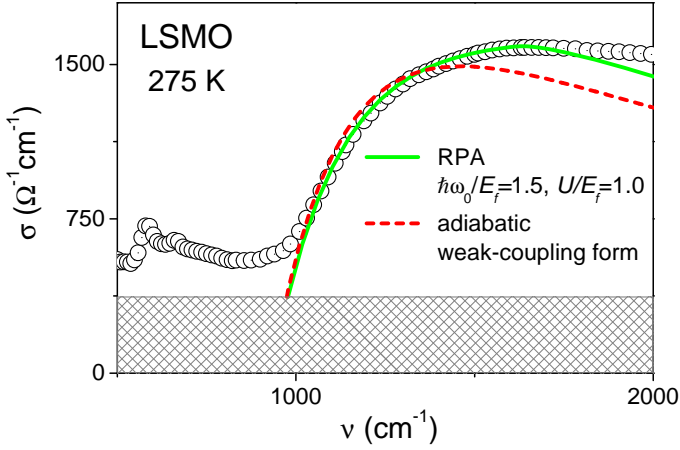}
\vspace{0mm} \caption[]{\label{fit-LP-Sr275K} (Color online) For
275~K: the RPA-approach, Eq.~(\ref{eq:HubbardSigma}), is evaluated
with a reduced Fermi energy (a reduction factor of 5 as compared
to the 100~K fit which is the continuous line in
Fig.~\ref{theory-Hubbard}). The adiabatic weak-coupling form,
Eq.~(\ref{eq:ScalingForm}), still coincides with the full
RPA-result and the experimental data along the low frequency
slope.}
\end{figure}

For higher temperatures, $\hbar \omega_0 /E_f \ll 1$ is not
necessarily true since $\omega_0(T)$ increases considerably (see
Table~\ref{TKFit}). Also the effective mass of the charge carriers
increases with temperature --- due to a reduced double exchange
--- which implies a reduced Fermi energy. For the 275~K data we
present in Fig.~\ref{fit-LP-Sr275K} the evaluation of
Eq.~(\ref{eq:HubbardSigma}) with a Fermi energy five times smaller
than for the low temperature fit. The dimensionless energy
parameters now take the values $\hbar \omega_0 /E_f\simeq 1.5$ and
$U /E_f\simeq 1$. Obviously, the RPA-result approximates the low
frequency slope of the polaronic resonance at 275~K with the same
rigor as for 6~K or 100~K. As anticipated, the adiabatic weak-coupling result
(Eq.~(\ref{eq:ScalingForm}), see dashed line in
Fig.~\ref{fit-LP-Sr275K}) now does not exactly match the full
evaluation but the deviations are still minor. To gain the
spectral weight of the measured polaronic resonance, the
electron-coupling has to be larger for the 275~K evaluation:
$\alpha\simeq 4-5$. This is to be expected since, with increasing
mass, the coupling should also be enhanced.\cite{comment3}
Although the weak coupling approach, Eq.~(\ref{eq:ScalingForm}),
is not fully applicable for this intermediate-coupling regime
at high temperatures, the result is still so convincing that we
propose to discuss all data sets, irrespective of temperature, within
the presented approach.

The curves for all considered temperatures have
the shape which is approximated by Eq.~(\ref{eq:ScalingForm})
for low frequencies, with the threshold $\omega_0$
replaced by $\omega_0(T)$. This statement agrees with the conjectured
scaling behavior for the low frequency slope, examined at the
beginning of the subsection on large polarons --- see Fig.~\ref{scaling},
in which the low-frequency behavior was actually extrapolated to the
``zero line'' with Eq.~(\ref{eq:ScalingForm}). The parameters for this
scaling form,  the amplitude $A(T)$ and the threshold frequency $\nu_0(T)$, are displayed in
Table~\ref{TKFit}.

\begin{table}[b]
\caption[]{$T$-dependent parameters of the approximate result
for LP, Eq.~(\ref{eq:ScalingForm}).} \label{TKFit}
\vspace{.2cm} \centering
\begin{ruledtabular}
\begin{tabular}{lccc cc cc cc cc cc cc}
  $T$ (K) && 6  && 100 && 160 && 200 && 250 && 275 && 295 \\
  \hline
  $\nu_{0}$ (cm$^{-1}$)&& 593 && 617 && 666 && 765 && 865 && 978 && 1183 \\
  $A$ (10$^{3} \, \Omega^{-1}$cm$^{-1}$)&& 18.3 && 16.3 && 14.3 && 12.4 && 9.5 && 7.9 && 6.4 \\
\end{tabular}
\end{ruledtabular}
\end{table}

Shortcomings of the weak coupling approach are the unexplained
(temperature dependent) threshold frequency $\omega_0(T)$ and the
unmatched slope at the high frequency side when we use a
reasonable parameter set in order to reproduce the spectral
weight. The coupling parameters are, with this restriction, in the
small coupling regime. The power law decay at the high energy side
is $\omega^{-2}$ in  Eq.~(\ref{eq:ScalingForm}) whereas the
observed power law is closer to $\omega^{-1}$ for the 6~K and
100~K data (and approximately $\omega^{-0.3}$ for the 275~K data).

The latter shortcoming should not be taken as a serious flaw of the
presented concept. The phase space for excitations increases strongly
when the excitation energy deviates from the proximity of the Fermi
energy and, moreover, the absorption through such ``high energy''
excitations is by no means universal. We cannot expect that the simple
treatment within RPA with quadratic dispersion can still capture these
high-energy excitations correctly.

The failure  of the weak coupling approach to account for the (temperature dependent)
threshold frequency $\omega_0(T)$ is more disturbing. The threshold in this approach is
fixed at the respective frequency of the lattice
deformation which will not shift from about 593~cm$^{-1}$
(threshold for 6~K) to  1183~cm$^{-1}$ (threshold for 295~K).
Resolving this issue  seems to require a type of
intermediate-coupling model where the threshold is renormalized
but the shape of the absorption spectrum is similar to the shape in the
weak coupling  approach. A possible scenario is that of
Kartheuser, Evrard and Devreese (KED, Ref.~\onlinecite{KED69};
see also the review by Mitra {\it et al.},\cite{Mitra87} section
2.6, for further references and for a discussion of  the
literature on this topic). KED propose that the optical absorption
for intermediate coupling consists of a relatively narrow
zero-phonon peak, assigned to a ``relaxed excited state'' (RES),
and of a broad (multi phonon) sideband, with threshold at
$\omega_{\rm LO}$ {\it above} the zero-phonon peak. Here,
$\omega_{\rm LO}$ is the frequency of the LO phonon. Consequently the
threshold will shift with the frequency of the RES which is
dependent on the electron-lattice coupling. This coupling hinges
on the electronic band width which in turn depends, for the
manganites, on temperature.
This scenario implies that a RES resonance exists at $\omega_{\rm LO}$ below the
polaronic continuum and that it shifts with temperature as the
threshold frequency $\omega_0(T)$ in our data.  However, such a
temperature dependent RES has not been observed  as an additional sharp
resonance for the manganites, neither in the optical conductivity
nor in Raman scattering. For this reason we consider the KED
scenario inapt to explain the temperature shift of the threshold in
the metallic LSMO films.

In spite of these shortcomings, it should be appreciated
that, with respect to the standard polaronic models, this TD-like approach
implements a Fermi surface. Short-range Coulomb interactions are
included within a controlled weak coupling scheme. Both,
Fermi statistics and Coulomb interaction, lead to substantial consequences for the
shape of the polaronic absorption spectrum.

As we already pointed out, a dynamical JT effect is missing in this
weak coupling approach. Mannella \textit{et al.}
investigated the temperature-dependent JT distortion (see Fig.~3 of
Ref.~\onlinecite{Mannella04}). The temperature dependence
of the variance of the Mn-O bond length distribution corresponds
with the temperature dependent shift of the polaronic MIR excitation
for both, LSMO and LCMO (see the inset of Fig.~2 in Ref.~\onlinecite{Hartinger04}).
It would be highly desirable to have an evaluation which parallels the
calculations of Millis, Mueller and
Shraiman~\cite{Millis96} for large polarons. For their local
modelling they find a temperature dependent shift of the polaronic
resonance for sufficiently strong coupling in the undoped case.
However, it is not yet conceivable how such a scheme can be set up for
extended excitations as the large polarons.

\subsubsection{Self-trapped polarons}
\label{sec:SelfTrapped}

The formation of large polarons in the manganites has been debated
before. R\"oder, Zang and Bishop\cite{Roeder96} studied  the
crossover from a large polaronic state to a quasi-self-trapped state
for a single hole in the adiabatic limit, applying a variational
technique for the DE model with the inclusion of one effective JT
mode. They used a mean-field evaluation for the spin degrees of
freedom and neglected other electronic correlation effects. The
optical absorption has not been considered in this approach.

Kim, Jung and Noh\cite{Kim98} propose that their optical
conductivity spectra of polycristalline LCMO
(La$_{0.7}$Ca$_{0.3}$MnO$_{3}$) display a sharp Drude peak and a
broad MIR absorption band which is to be explained as coherent
tunnelling and an incoherent band of large lattice polarons,
respectively. They particularly emphasize that, with their
electrodynamic analysis, the large polaron should be in a strong
coupling regime --- a polaronic state which indeed has not been
realized in other physical systems because the polaron would
usually collapse to a small polaron state for strong coupling.
However, it is not obvious how to identify  the {\it large}
polarons from their optical conductivity data. A fit to the large
polaron picture by Emin\cite{Emin93}, the model to which the
authors of Ref.~\onlinecite{Kim98} refer to, has not been
presented. The steep low-frequency slope of this type of polaronic
absorption cannot be resolved in their data. However, since in our
thin film optical spectra of LSMO we can clearly identify such a
threshold behavior, we will now reinvestigate Emin's
phenomenological approach in some detail.

The absorption process arises from exciting a carrier from the
ground state in its (self-trapping) potential well to a
free-carrier state, invoking the Franck-Condon principle. In this
intuitive approach, neither the mechanism which is responsible for
the self-trapping nor the nature of the extended states are
specified. The optical absorption is calculated by applying
Fermi's Golden Rule\cite{Emin93} which requires the identification
of the initial and final charge carrier state. Here we slightly
deviate from Emin's evaluation by picking a Gaussian localized
state $|G\rangle$ instead of a hydrogenic ground state as initial
state. The polaron energy of $|G\rangle$ is lower and it is better
confined (see, for example, Ref.~\onlinecite{Mitra87}). The
``Gaussian'' result for $\sigma(\omega)$ will deviate not
significantly from that for the hydrogenic state. The initial
localized electronic wave function is $\langle
r|G\rangle=(1/R\sqrt{\pi})^{3/2}\,\exp(-r^2/2R^2)$, and the final
free-carrier wave function is $\langle {\bf r}|{\bf k}\rangle=
L^{-3/2}\exp(i{\bf kr})$. Emin assumes free-carrier states with
quadratic dispersion $\varepsilon=k^2/2m$ which, for the optical
absorption, translates into a frequency dependent wave vector:
$k(\omega)=\sqrt{2m(\hbar\omega-E_0)}/\hbar$, where $E_0/\hbar$ is
the threshold frequency for the absorption ($E_0$ is the energy
difference between the localized ground state and the lowest
continuum state). The optical conductivity is then:
\begin{equation}
\label{eq:GaussianPolaron}
\sigma(\omega)\;=\;
n_p \frac{8\sqrt{\pi}}{3}
\frac{e^2}{m}\;\frac{1}{\omega}\;\bigl(k(\omega)R\bigr)^3
\;e^{-\bigl(k(\omega)R\bigr)^2}
\end{equation}
The best fit for the 100~K data is displayed in Fig~\ref{theory}.
If we assume that $m=3m_e$, consistent with specific
heat measurements,\cite{Salamon01} and $E_0/h=543$~cm$^{-1}$ we find a
polaron radius from this fit of approximately $R/a\simeq
2.15/\sqrt{3}$, where $a\simeq3.88$~\AA. Correspondingly, the
diameter of the polaronic (ground) state is about 2.5 lattice
constants (for $m=m_e$, $R/a\simeq 2.15$ the diameter is more than
4 lattice constants). The maximum height of the resonance fixes
the prefactor of Eq.~(\ref{eq:GaussianPolaron}): the polaron
density is estimated for $m=3m_e$ to a value of $n_p=3 \times
10^{21}$~cm$^{-3}$ (and for $m=m_e$ to $n_p=10^{21}$~cm$^{-3}$).
The values for $R$ and $n_p$ are reasonable in this context and
support this approach. We point out that the hydrogenic ground
state, as introduced by Emin and considered in our previous work,
generates a polaron radius which is already unacceptable
small.\cite{Hartinger04}

Next we discuss the scaling behavior introduced above (see
Fig.~\ref{scaling}, upper panel). If we rescale $\sigma(\omega)$ and $\omega$
as
\begin{equation}
\label{eq:ScalingRelations}
\hat{\sigma}\;\equiv\;
\sigma\,\bigg/\,
     \biggl(
       \frac{8\sqrt{\pi}}{3} \frac{n_p}{m} \frac{e^2\hbar}{E_0}
    \biggr) \qquad {\rm and} \qquad
w \equiv \frac{\hbar\omega}{E_0},
\end{equation}
respectively, and if we also introduce a dimensionless radius
\begin{equation}
\label{eq:Radius} \varrho\; \equiv\; R\, \sqrt{2mE_0}/\hbar \, ,
\end{equation}
 we identify the rescaled optical conductivity as
\begin{equation}
\label{eq:GaussianScaling} \hat{\sigma}(w)\;=\; \frac{1}{w
}\;\bigl(\varrho\,\sqrt{w-1}\bigr)^3 \;\exp
\bigl(-\varrho^2\,(w-1)\bigr) \, ,
\end{equation}
which only depends on the dimensionless radius $\varrho$.
If $\varrho$ is independent of temperature, the observed
scaling is also true for this approach. For the 100~K data,
$\varrho = 1.16\simeq\sqrt{3/2}$. The fits for the other
temperatures produce $\varrho = 1.10\, ... \,1.22$ (see Table~\ref{GaussFit}),
a range which is sufficiently narrow to support the observed scaling within
this approach.

\begin{table}[b]
\caption[]{Self-trapped LP:  parameters from
Eqs.~\ref{eq:GaussianPolaron}--\ref{eq:GaussianScaling} to fit the
LSMO MIR spectra at different temperatures for $m=3m_{e}$.}
\label{GaussFit} \vspace{.2cm} \centering
\begin{ruledtabular}
\begin{tabular}{llcc cc cc cc cc cc cc}
  $T$ (K) &&& 6 && 100 && 160 && 200 && 250 && 275 && 295 \\
  \hline
  $E_{0}$ (cm$^{-1}$)&&& 526 && 543 && 580 && 650 && 777 && 876 && 1060 \\
  $\varrho$ &&& 1.22 && 1.17 && 1.13 && 1.15 && 1.14 && 1.10 && 1.15 \\
  $R/a$ &&& 1.96 && 1.86 && 1.73 && 1.67 && 1.51 && 1.37 && 1.31 \\
  $n_{P}$ (10$^{21}$cm$^{-3}$)&&& 2.8 && 2.7 && 2.6 && 2.4 && 2.34 && 2.26 && 2.1 \\
\end{tabular}
\end{ruledtabular}
\end{table}

Even so, is the experimentally estimated value of $\varrho\simeq\sqrt{3/2}$
supported by theoretical considerations?
In the Landau-Pekar strong coupling theory
with a Gaussian trial wave function (see, for example,
Ref.~\onlinecite{MahanBook}) the energy of the localized polaron
is $E_p=-(\alpha^2/3\pi)\,\hbar \omega_{\rm LO}$ and
$R=(3\sqrt{\pi/2}/\alpha)\;\ell$, with the phononic length scale $\ell \equiv
\sqrt{\hbar/(2m\omega_{\rm LO})}$. Correspondingly, the polaron
binding energy may be reexpressed in terms of the polaron radius:
 $E_p=-(3/2)\,(\hbar^2/(2mR^2))$. If we assumed that $|E_p|=E_0$ holds
and if we use the above relation between $R$ and $\varrho$, we
would confirm exactly  $\varrho=\sqrt{3/2}$. Then, since
$\alpha=\sqrt{3\pi}\sqrt{E_0/\hbar\omega_{\rm LO}}$, we would have
for the 6~K data $\sqrt{E_0/\hbar\omega_{\rm LO}}\simeq 1\, ...
\,1.4$ and correspondingly $\alpha=3\, ... \, 4$, and for the
295~K data $\sqrt{E_0/\hbar\omega_{\rm LO}}\simeq 1.4\, ... \,2$
with  $\alpha=4\, ... \, 6$. Although this signals the
intermediate coupling regime, with non-negligible further
corrections to the polaron binding energy, the estimates seem to
be in a reasonable range.

However, a serious objection is to be raised against this
procedure: $E_0$ can be identified with $|E_p|$ only if the
self-trapped state disappears together with the polarization well
in the excitation process. This should not be the case in the
considered approach in which we investigate excitations in the
adiabatic limit and invoke the Franck-Condon principle. In this
limit, we should take $E_0= 3|E_p|$ (see Emin's
review,\cite{Emin93} or the book by Alexandrov and
Mott,\cite{AMBook} chapter~1.1). With $E_0= 3|E_p|$ we find
$\varrho=3/\sqrt{2}$ which does not fit the experimental
data.\cite{comment4} Next order corrections to the Landau-Pekar
limit do not alleviate this inconsistency.

{\it Assessment of the strong coupling scenario.} The resonance in
the MIR optical conductivity can be well fitted with Emin's
phenomenological strong-coupling approach. Again, the fit is
convincing up to frequencies slightly above the maximum whereas
the high frequency slope decreases too fast, even faster than for
the weak coupling scenario. The observed scaling is approximately
reproduced, however with a value for the dimensionless radius
$\varrho$ which is inconsistent with a strong coupling estimate.

It is mandatory to ask if this phenomenological approach is trustworthy
in terms of microscopic considerations. Alexandrov, Kabanov and
Ray\cite{Alexandrov94} presented numerical results for the Holstein
model with two and four sites and up to 50 excited phonons per
mode. In fact, the shape of this microscopically calculated polaronic
absorption resonance is, for intermediate coupling, similar to that of the
phenomenological approach: a steep slope above the threshold and
slower decrease at the high frequency side, comparable
to the fit curve (Gaussian polaron) in  Fig.~\ref{theory}. A direct
confirmation of the phenomenological approach was attained for the
one-dimensional model.\cite{Alexandrov94} Hence we consider
Emin's approach to be reliable for the considered adiabatic limit.

Despite this positive assessment of the theoretical consistency of
the approach we have to question its validity for the metallic
manganites. The material (LSMO) is conducting and the effective
mass of the charge carriers is supposed to be enhanced but not
extremely strong. Consequently, a Fermi surface of the carriers
should exist above the mobility edge, at least for low
temperatures (we do not explore any details here; for a more
comprehensive discussion see the review article by
Edwards\cite{Edwards02} and the literature cited there, especially
Refs.~\onlinecite{Livesay99},~\onlinecite{Pickett97}). This
implies that if such Gaussian localized polarons exist they have
to be excited across the mobility edge to the Fermi surface ---
not to a band minimum as assumed in the above polaron
model.\cite{comment4a}

The consequence now is that the phase space for the excitation of
a trapped polaron into ``free carrier states'' at the hole Fermi
surface is large, the density of states is finite, and one expects
a step-like increase of the absorption. The experimental shape of
the polaronic absorption is, however, not consistent with a
(possibly smoothed) absorption-edge step.

Furthermore, one has to account for a mobile charge carrier
density of the order of $n=10^{21}$~cm$^{-3}$ to
$10^{22}$~cm$^{-3}$. This would push the mobility
edge\cite{comment4b} to energies well below the Fermi edge with
the consequence that the frequency of the absorption edge is
certainly larger than the observed threshold --- at least for the
low temperature measurements. With this in mind one should dismiss
the scenario that all the polarons which contribute to the
absorption are self-trapped, that is, localized. They may have a
large polaronic mass but they are mostly delocalized. However,
Emin's approach decisively depends on localized initial
states.\cite{comment5} For a transition from a band state to
another free-carrier state one again expects a sharp increase of
the absorption.\cite{comment6}  With these impediments it is
unlikely that this intuitive approach can be applied directly to
the observed polaronic absorption in the LSMO films.

\subsubsection{Fermi edge singularities}
\label{sec:FermiEdge}

A generalization of the previous considerations (in
Sec.~\ref{sec:SelfTrapped}) to a finite concentration of mobile
carrier states is necessary but its implementation is subtle:
In the adiabatic limit the localized potential well, formed by the
lattice distortion, acts after the excitation of
the charge carrier to the Fermi edge as a potential scatterer.
Consequently, this limit represents the x-ray edge problem and
the power law above the absorption threshold depends on the values of the
scattering phase shifts.\cite{Nozieres69} It may diverge (the
Mahan exciton phenomenon is dominant) or converge (orthogonality
catastrophe). Since we have to expect many decay channels for this
excitation, the singular diverging edge might well be smoothed so
as to reproduce the shape observed in the measurements. The decay
at the high energy slope is governed by $\alpha_l =
2\,\delta_l(k_f)/\pi-\alpha$ where $\delta_l(k_f)$ is the phase
shift in the angular momentum channel $l$ and $\alpha$ is the sum
over all phase shifts squared (see, for example,
Ref.~\onlinecite{MahanBook}). The absorption  process involves,
according to the standard selection rules, power law behavior with
both exponents, $\alpha_{l\pm 1}$. However it is not evident, why
$\alpha_{l\pm 1}$ should be so strongly temperature dependent as
to reduce the exponent in the power law to the appropriate
($T$-dependent) value. Nevertheless, $\alpha$ might well increase
with more scattering channels available when approaching a
disordered paramagnetic state with increasing temperature.

Above, we argued that we do not expect the polarons, which
participate in the absorption within the considered frequency
window, to be localized. With finite polaron dispersion the
situation changes qualitatively\cite{Gavoret69,Ruckenstein87}
because the recoil of the polaronic potential well has to be
accounted for.  The edge singularity will form a resonance around
the frequency of the direct transitions from the filled valence
band to the Fermi edge in the conduction band. This resonance
(Mahan exciton) will flatten out when the ratio of the valence
band mass to the conduction band mass increases. It is not clear
how to identify the polarons with a valence band --- or how to
quantify a ``valence band hole'' as the appropriate potential well
of the lattice distortion stripped of its charge carrier. In spite
of that, the mechanism discussed by Gavoret, Nozi\`eres, Roulet
and Combescot\cite{Gavoret69}  and later on by Ruckenstein and
Schmitt-Rink\cite{Ruckenstein87} is still effective even if the
large polarons may not be visualized as forming a valence band of
quasi particles. It is the collective behavior associated with the
response of the Fermi sea in the course of the excitation process
that determines the optical spectrum. However, we have no
information on the frequency difference of direct to indirect
absorption and we have neither any information if the effective
mass of the free carriers at the Fermi surface increases with
respect to that of the delocalized polaronic potential well in order to
produce the observed temperature dependence of the spectra. To
summarize, in a scenario of adiabatic large polarons we have to
expect these collective effects of the Fermi sea but it is
speculative if they dominate the shape of the optical spectra in
the considered frequency window. Furthermore, deviations from the
adiabatic limit will contribute to a smoothing of the edge
singularity.

\section{Conclusions}
We have presented experimental results of the optical conductivity
and dc-resistivity in thin films of LSMO and LCMO to analyze
polaronic excitations. In both cases the polaron formation energy
increases with  temperature up to $T_{\rm MI}$, while the spectral
weight diminishes continuously.

Our data indicate different polaronic regimes in both compounds. For
LCMO, the SP picture is well accepted, particularly with regard to
the high temperature dc-resistivity. The optical conductivity
displays an asymmetric Gaussian line shape close to the maximum of
the polaronic resonance which also suggests incoherent tunnelling
of small polarons. Overall, we conclude that for the Ca-doped
compound the SP model with thermally activated transport is
approximately valid for both, the ferromagnetic as well as the
paramagnetic phase.

In contrast, for LSMO, ln($\rho$/T) deviates from a linear behavior
and the optical conductivity has a strong asymmetric line shape with
a steep slope at the low frequency side. This shape is typical for
LP absorption processes.\cite{MahanBook,Emin93,Tempere01} The
phenomenological adiabatic LP model by Emin\cite{Emin93} considers
excitations of self-trapped charge carriers to an empty band of
free-carrier states, with the polaronic potential-well intact in the
adiabatic limit. Emin's model reproduces the steep low frequency
slope with a sufficiently wide maximum but cannot account for the
slow decrease on the high frequency side. We challenge that this
approach is applicable here because there is no hole-band minimum
available to which the excited charge carriers could scatter. The
energy difference from the mobility edge, below which the trapped
polarons would reside, to the hole band minimum would have to be the
threshold frequency. This frequency $E_0/h$ is (for 6~K) about
$526$~cm$^{-1}$, too small for the addressed energy difference.
Consequently the modelling has to be modified as to implement LPs
with finite mass and a finite density of charge carriers (Fermi
edge).

In the weak coupling theory, the generalization of the single
polaron absorption by GLF\cite{Gurevich62} to the situation of
many-particle absorption was carried out by Tempere and
Devreese.\cite{Tempere01} The optical conductivity for a similar
approach, which is rather based on the Hubbard model than on a
Coulomb gas ansatz, is presented in Figs.~\ref{theory},
\ref{theory-Hubbard}, and \ref{fit-LP-Sr275K}. The high frequency
decay is now better realized through the excitation of particle-hole
pairs of the Fermi sea. The scaling of the low-frequency slope (as
observed in Fig.~\ref{scaling}, upper panel) is implemented through
the increase of the number of particle-hole pairs with increasing
energy from the Fermi edge. On the other hand, this model cannot
account for the shift of the threshold with temperature; the model
threshold would always be the LO phonon frequency which is not
supposed to shift from $593$~cm$^{-1}$ for 6~K to $1183$~cm$^{-1}$ for 295~K.

The failure of weak coupling theory
necessitates to extend these considerations to intermediate or
strong coupling situations, first and foremost in the adiabatic
limit. Such an evaluation is still not available and hardly
feasible. As we discussed, it involves a collective response of the
Fermi sea which leads to singularities in the optical conductivity,
well known from the x-ray edge problem. The recoil of the polaronic
lattice deformation, which is taken up in the excitation process of
the charge carrier to the Fermi edge, will flatten the singularity
and might be responsible for the observed shape. Although the
proposed consequences for the shape of the absorption spectrum are
speculative, the singular response is supposed to be present in any
model with a sudden creation of a heavy or localized scatterer in
the presence of a Fermi sea.

To summarize, the comparison between the measured optical conductivity
spectra and the current theoretical models exhibits their inadequacy
to explain the spectral weight at the high frequency slope
of the polaronic MIR resonance and the strong shift of the
resonance position with temperature.
A theoretical solution of the polaron problem which may hold for
any electron-phonon coupling, including correlation effects, is
highly desirable. The physical quantities that control the
temperature dependence for polaron formation, still remain to be
elucidated.

Thin films proved to be superior for the analysis of the
phononic\cite{Hartinger04c1,Hartinger04c2} and polaronic spectra in
metallic manganites. We expect that further investigations based on
thin film samples will provide considerable additional insight into
the physics of the metallic state of the CMR manganites. We refer to
an examination of the doping dependence of
La$_{1-x}$Sr$_{x}$MnO$_{3}$, the polaronic spectra of which may
address the built up of the Fermi surface. We also propose to
explore the magnetic field dependence of the thin film spectra in
order to address the role of magnetic correlations in the formation
of polarons. Finally,
La$_{0.75}$(Ca$_{0.45}$Sr$_{0.55}$)$_{0.25}$MnO$_{3}$ undergoes an
orthorhombic to rhombohedral transition with increasing
temperature\cite{Kiryukhin03} which would invite to examine a
possible crossover from SP to LP behavior in the MIR optical
conductivity.

\begin{acknowledgments}
We acknowledge discussions with R.~Hackl and H.-A.~Krug von Nidda.
X-ray measurements were performed at the Walther-Meissner-Institut
(Garching). We thank A.~Erb und P.~Majewski for
assisting the x-ray measurements. This research was supported by the
BMBF (13N6917 and 13N6918) and by the Deutsche Forschungsgemeinschaft
through SFB 484 (Augsburg).
\end{acknowledgments}



\begin{thebibliography}{1}
\bibitem[{com({\natexlab{a}})}]{comment0}
\bibinfo{note}{{The LSMO sample of reference~\onlinecite{Hartinger04} is not
  identical to the LSMO sample presented in this paper. For this reason the
  transition temperatures slightly differ. However the optical spectra are so
  similar that our analysis applies to both samples without reservation.}}
\bibitem[{com({\natexlab{b}})}]{comment1}
\bibinfo{note}{{The coherent motion from quantum tunneling with a fixed number
  of phonons generates a polaron band with bandwidth that decreases
  exponentially with temperature --- for temperature larger than the Debye
  scale. We do not attempt to identify this polaron band with the low-frequency
  absorption upturn. Certainly, other mechanisms have to be incorporated to
  fully deal with the low-energy IR-absorption (see, for example,
  Refs.~\onlinecite{Ishihara97, Mack99}).}}
\bibitem[{com({\natexlab{l}})}]{commentCPA}
\bibinfo{note}{{A discussion and the relevant literature on this CPA-approach
  is presented in Ref.~\onlinecite{Edwards02}}}.
\bibitem[{com({\natexlab{d}})}]{comment3}
\bibinfo{note}{{In the Fr\"{o}hlich Hamiltonian, the electron-phonon coupling
  is $\alpha=(e^2/\hbar)(m/(2\hbar\omega_0))^{\frac{1}{2}}(1/\epsilon_\infty-1/
  \epsilon_0)$. With $\hbar\omega_0\simeq 10^{-20}$~J,
  $m=3m_e$, $\epsilon_\infty\simeq4$ and $\epsilon_0\simeq 16$, we would find
  $\alpha\simeq 4.5$. A value of 3 to 4 for $\alpha$ is consistent with the
  literature.}}
\bibitem[{com({\natexlab{e}})}]{comment3a}
\bibinfo{note}{{The weak coupling limit, for which neither shape nor weight
  change noticeable, is approached for $U/E_{f} \lesssim 0.01$.}}
\bibitem[{com({\natexlab{g}})}]{comment4}
\bibinfo{note}{{A similar discussion holds for the hydrogenic ground state of
  the polaron with $\psi(r)=1/\sqrt{\pi R^3}\exp(-r/R)$, except that the
  parameter values differ. For example, $E_p=- \hbar^2/(2m R^2)$ which implies
  $\varrho =1$ for that case, assuming that $E_0=|E_p|$ holds (it is $\varrho
  =\sqrt{3}$ for $E_0=3|E_p|$). The analytical form of the optical absorption
  is found in Emin's review. The high frequency decrease is slightly slower in
  the frequency range of sizable absorption but it does not improve the fit to
  the experimental data.}}
\bibitem[{com({\natexlab{h}})}]{comment4a}
\bibinfo{note}{{It would be unrealistic to assume that the bound polarons are
  particles from the electron spheroids and that they are excited back to the
  empty electron band centered at the $\Gamma$ point, a situation which would
  be compatible with the above approach. To account for the large spectral
  weight one rather has to suppose that it is the hole carriers which form the
  (trapped) polarons and which are excited to the Fermi surface of the hole
  cuboids (at the $R$ $(\pm \pi/a,\pm \pi/a \pm \pi/a)$ points --- other empty
  hole bands are not sufficiently low in energy). The hole nature of the
  polaronic charge carriers also agrees with the standard microscopic
  considerations.}}
\bibitem[{com({\natexlab{i}})}]{comment4b}
\bibinfo{note}{{Assuming that a mobility edge exists for a many-polaron system
  --- there is no strict proof yet.}}
\bibitem[{com({\natexlab{j}})}]{comment5}
\bibinfo{note}{{The localized state can take up any momentum from the
  excitation.}}
\bibitem[{com({\natexlab{k}})}]{comment6}
\bibinfo{note}{{The theory with final state interactions for Wannier excitons
  predicts a step-like increase when the excited bound states are disregarded;
  see Ref.~\onlinecite{MahanBook}, section 9.2.}}
\end{thebibliography}

\end{document}